\documentstyle[preprint,aps]{revtex}
\begin{document}
\draft
\preprint{NSF-ITP-93-43}
\title{Subharmonic generation in quantum systems}
\author{Martin Holthaus}
\address{Department of Physics, Center for Nonlinear Science, and
Center for Free-Electron Laser Studies, \\
University of California, Santa Barbara, CA 93106-9530}
\author{Michael E. Flatt\'e}
\address{Institute for Theoretical Physics, University of California,
Santa Barbara, CA 93106-4030}
\date{April 19, 1993}
\maketitle
\begin{abstract}
We show how the classical-quantum correspondence permits long-lived
subharmonic motion in a quantum system driven
by a periodic force. Exponentially small deviations from
exact subharmonicity are due to
coherent tunneling between quantized vortex tubes
which surround classical elliptic periodic orbits.
\end{abstract}

\narrowtext

When atoms interact with intense laser fields, extremely high harmonics can
be generated.  In recent experiments subjecting Ne atoms to
1-ps $10^{15}$-W/cm$^{2}$ laser pulses, harmonics up to the 135th were
detected~\cite{EXP}.

The question then arises whether nonlinearities in the
laser-matter interaction can also lead to the generation of
subharmonics. Subharmonic motion occurs quite naturally in classical
periodically-forced nonlinear oscillators, and recent investigations
have shown that the correspondence principle extends
much farther than
previously believed~\cite{GUT,BEK,WIN,STE,HEL,BTU}.
The purpose of
this Letter is to demonstrate how classical subharmonic motion survives in the
corresponding quantum system.

To illustrate our considerations, we use the model of
a particle in a one-dimensional triangular
well interacting with
an external periodic force\cite{SMI,CAS}.
We employ dimensionless ``atomic units''
such that the particle mass, the potential slope, and Planck's quantum
$\hbar$ are unity, and denote the strength and frequency of the driving
force by $\lambda$ and $\omega$. The Hamiltonian is then given by
\begin{equation}
H(p,x,t) = \frac{p^{2}}{2} + x + \lambda x \sin(\omega t)\label{HAM}
\end{equation}
for $ x \ge 0 $, with a hard wall at $x = 0$.

Fig.~1 shows a Poincar\'{e} surface of section for the classical version
of this system with
$ \lambda= 0.4 $ and $\omega = 0.92$, plotted in action-angle
variables $(I,\varphi)$ of the
undriven well. In the lower left corner, the elliptic island that originates
from the primary 1:1 - resonance is visible. This island is organized
around a stable (elliptic) periodic orbit which closes on itself after one
period $T$ of the external force. In contrast to the chaotic motion of
trajectories in the surrounding stochastic sea, motion inside this
island is mainly regular. A major fraction of the extended phase space
$ \{(p,x,t)\}$ surrounding the periodic orbit is filled with invariant
$T$-periodic vortex tubes~\cite{ARN}, and a trajectory with an initial
condition on such a tube remains confined to it for all times $t$.
The closed curves inside the islands seen in Fig.~1 are sections of vortex
tubes with the plane $ t = 0 $.

Disregarding small secondary resonances, the main other features of the
Poincar\'{e} section are the two large islands which belong to the
2:1 - resonance. In this case, the central periodic orbit closes
only after two cycles of the driving field, and the surrounding
vortex tubes are $2T$-periodic. The appearance of two islands
in Fig.~1 is simply a consequence of the fact that the classical flow
is sampled once every period, or, expressed differently, that the
projection of the $2T$-periodic vortex tubes to the fundamental
part of phase space
$ \{(p,x,t) , 0 \leq t < T\} $
consists of two
disconnected pieces. Thus, ``inside'' the 2:1 - resonance there is a portion
of phase space that sloshes with half the frequency of the external drive.

If the area of these islands is large enough to support quantum
states, this subharmonic
sloshing should have consequences for the corresponding
quantum system.
The scale for the classical-quantum comparison is set by Planck's
constant. Since $\hbar=1$ in our units, the action
$I_n = n+3/4$ (starting from $n=0$) corresponds semiclassically to the
$n$-th quantum state. Thus Fig. 1 indicates that the 2:1 -
resonance will strongly influence quantum states between (roughly)
$n=25$ and $n=50$.

In general, the quantum dynamics of a system with a $T$-periodic
Hamiltonian operator $ H(p,x,t) $ should be investigated in terms of its
Floquet states $u(x,t)$, i.e.~the $T$-periodic eigensolutions of the
equation
\begin{equation}
\left( H(p,x,t) - i\partial_{t} \right) u(x,t) = \varepsilon u(x,t)
\;\;\; .
\end{equation}
The Floquet states form a complete set in an extended Hilbert space
of square integrable, $T$-periodic functions~\cite{SAM}. Their role
in periodic systems is analogous to that of energy eigenstates in
time-independent systems. The eigenvalues $\varepsilon$, which are
called quasienergies~\cite{ZEL,RIT}, are defined up to an integer
multiple of $  \omega $ and can therefore be arranged in
Brillouin zones, with the first zone ranging from $ -\omega/2 $
to $ +\omega/2 $.

Quasienergies for the driven triangular well are shown in Fig.~2
as functions of $ \lambda $. Only those quasienergies
that connect for $ \lambda\rightarrow 0 $ to the energies of the unperturbed
stationary states from $ n = 20 $ to $ n = 50 $ have been displayed.
The influence of the classical 2:1 - resonance is clearly visible:
There are two ``fans'' of quasienergies that appear to behave almost
identically with coupling strength, except that they differ by an amount
very close to $ \omega/2$ (mod $\omega$).

It is important to realize that such a spectral feature is not due to
the specific triangular well potential, but solely to the presence
of a resonance.
Let $ \chi_{n}(x) $ denote the energy eigenfunctions of the undriven
well, and $ E_{n} $ their energies. The existence of a classical
2:1 - resonance corresponds, in the quantum system, to an energy
spacing which is close to $ \omega/2 $ in the vicinity of a certain
state $ n_{0} $. Thus, we also assume $ E'_{n_{0}} = \omega/2 $
(the prime denotes differentiation with respect to quantum number)
and expand the wave function as
\begin{equation}
\psi(x,t) = \sum_{n} c_{n}(t) \chi_{n}(x) \exp\left\{-i(n\omega/2
+ E_{n_{0}})t \right\}	\;\;\; .
\end{equation}
If only resonant terms are kept, one obtains for the coefficients
$ c_{n}(t) $ the equation
\begin{equation}
i\dot{c}_n =
\left(E_{n} - n\frac{\omega}{2} - E_{n_{0}} \right) c_{n}
+ V_{n,n+2}c_{n+2} + V_{n,n-2}c_{n-2}
\end{equation}
with $ V_{n,m} = \lambda \langle n | x | m \rangle / 2 $.
This equation leads to two separate groups of states, since it decouples
coefficients with odd from those with even indices.
Expanding the energies $ E_{n} $ quadratically around
$ E_{n_{0}} $ and assuming $ V_{n,n+2} = V_{n,n-2} \equiv V_{0} $
to be a constant, standard techniques~\cite{BER,LON} can be employed to
express the resonant Floquet states in terms of Mathieu functions~\cite{ABR}.
For even $ n-n_{0} $, the quasienergies are found to be
\begin{equation}
\varepsilon_{k}(q) = E_{n_{0}} + \frac{1}{2} E''_{n_{0}} \alpha_{k}(q)
\;\;\; \bmod \omega
\end{equation}
where $ q = 2V_{0}/E''_{n_{0}} $, and $ \alpha_{k}(q) $ is a characteristic
value of the Mathieu equation that is associated with a $\pi$-periodic
Mathieu function, i.e.~one of those characteristic values usually
denoted by $ a_{0}, b_{2}, a_{2}, \ldots $~\cite{ABR}.
If, however, $ n - n_{0} $ is odd, the quasienergies are
\begin{equation}
\varepsilon_{k}(q) = E_{n_{0}} + \frac{1}{2} E''_{n_{0}} \alpha_{k}(q)
+ \frac{1}{2}\omega \;\;\; \bmod \omega \;\;\; ,
\end{equation}
where now $ \alpha_{k}(q) $ is one of the characteristic values
$ b_{1}, a_{3}, b_{3}, \ldots $, belonging to a $2\pi$-periodic Mathieu
function. Despite the approximations, these formulae
yield a quite good description of the numerically computed
spectrum~\cite{LON}. Most importantly, the known asymptotic behavior
of the characteristic values~\cite{ABR} leads to an an~alytical estimate
for the quasienergies in the strong driving regime. For instance,
the difference between the two quasienergies indicated by the
arrows in Fig.~2 is given, modulo $\omega$, by
\begin{equation}
\frac{\omega}{2} +
\frac{1}{2}E''_{n_{0}}\left(b_{1} - a_{0} \right) \approx
\frac{\omega}{2}+
E''_{n_{0}} \, 2^{4} \sqrt{\frac{2}{\pi}} \, q^{3/4} \exp(-4\sqrt{q})
\; . \label{SPL}
\end{equation}
Since $q$ is proportional to $ \lambda $, we find that the deviation
from $ \omega/2 $ becomes exponentially small with the square root of
the coupling strength. For $ \lambda = 0.4 $, we estimate this deviation to be
$ 4 \cdot 10^{-18} \, \omega $.

A coherent superposition of two Floquet states $u_{1}(x,t)$ and $u_{2}(x,t)$
with quasienergies $ \varepsilon_{1} $ and $ \varepsilon_{2} = \varepsilon_{1}
+ \omega/2 $,
\begin{equation}
\psi = A_{1}u_{1}\exp\{-i\varepsilon_{1}t\}
          + A_{2}u_{2}\exp\{-i(\varepsilon_{1} + \omega/2)t\},
	\label{WAV}
\end{equation}
will radiate at half the driving frequency. This occurs because the
dipole $\langle\psi | x | \psi\rangle$ is not $T$-periodic --- its
shortest cycle time is $2T$. A spectrum like that in Fig.~2 thus
suggests
the possibility of subharmonic generation which, neglecting the
exponentially small corrections~(\ref{SPL}), is independent of
the precise value of the driving amplitude $\lambda$. This possibility
survives even for pulses with a slowly varying amplitude, to which the wave
function~(\ref{WAV}) can respond adiabatically.

We now interpret these results from a semiclassical point
of view. Floquet states and quasienergies can be calculated approximately
by means of semiclassical quantization rules which are similar to the
Einstein-Brillouin-Keller (EBK) conditions~\cite{BRE,KOR}. For driven
one-dimensional systems, like the particle in a triangular well, they
can be written as
\begin{equation}
\oint_{\gamma_{1}} \! pdx = 2\pi \left(n_{1} + \frac{1}{2} \right)
\;\;\; ,        \label{RU1}
\end{equation}
where the quantization path $ \gamma_{1} $ winds around a
$T$-periodic vortex tube in a plane of constant time $t$, and
\begin{equation}
\varepsilon = -\frac{1}{T} \int_{\gamma_{2}} \! (pdx - Hdt)
+ n_{2}\omega      \label{RU2}
\end{equation}
with a $T$-periodic path $ \gamma_{2} $ that lies on such a tube,
and the integration extending over one period.
The integer $ n_{1} $ is the semiclassical quantum number;
$ n_{2} $ accounts for the mod $\omega$ multiplicity of
the quasienergies $\varepsilon$. It is crucial to realize that the application
of these rules requires the existence of $T$-periodic vortex tubes.
They can be applied to vortex tubes that are merely perturbative (and,
therefore, $T$-periodic) deformations of energy manifolds of
the undriven system, or to the tubes of a 1:1 - resonance, but {\em{not}}
to any other resonance. The first rule~(\ref{RU1}) will always allow the
selection of quantized vortex tubes and the construction of associated
semiclassical wave functions, but the $T$-periodic boundary
conditions of the second one~(\ref{RU2}) are incompatible with,
e.g., the $2T$-periodic tubes of a 2:1 - resonance, so that these functions
are no Floquet states.

Such a contradiction can only be resolved if a Floquet state is not associated
with a single quantized $2T$-vortex tube, but with both parts that result
from its projection to the fundamental time interval. In this way,
$T$-periodicity is restored. Because there are two equivalent quantized
$2T$-tubes, the Floquet states must appear in pairs, with almost the
same probability density for both members of such a pair.

This reasoning is confirmed by a numerical computation of the exact Floquet
states. Fig.~3 shows, for $ \lambda = 0.4 $, the probability density of one
of the two states that belong to the marked quasienergies in Fig.~2.
The density is concentrated along {\em{both}} realizations of the periodic
orbit that bounces against the wall at $x=0$ and is reflected.
As expected, the density of
the corresponding state in the second quasienergy fan is almost the same.

Conceptually, the situation encountered here is strongly
reminiscent of a particle in a double well~\cite{LON}. An attempt to
calculate the eigenstates in a double well potential by applying the simple
Bohr-Sommerfeld quantization separately to the two wells, without accounting
for tunneling through the barrier, yields states which are strictly confined
to the individual wells. The correct eigenstates appear in doublets
that are delocalized over both wells (odd and even combinations).
Analogously, the Floquet state shown
in Fig.~3 is a member of the ground state doublet of the 2:1 - resonance;
the other corresponding states in the two fans constitute the excited
doublets. This also explains the $\omega/2$ difference in
quasienergies between states in a doublet. After one period $T$, the
phases of the odd and even combinations differ by $\pi$.

In an ordinary double well, localized states can be realized by appropriate
linear combinations of the eigenstates. In the same way, linear combinations
of a Floquet-doublet lead to a density which is concentrated asymmetrically
along only one of the two equivalent tubes. Such a density is $2T$-periodic,
and that is why a superposition like~(\ref{WAV}) generates subharmonics.
An example of a superposition of the 2:1 ground state doublet is shown in
Fig.~4; the corresponding expectation value of the dipole operator
is plotted in Fig.~5 for an interval of six periods
of the driving force. It is obvious that this expectation value contains a
strong subharmonic contribution; a Fourier analysis confirms that the
subharmonic mode is by far the dominant one.

If there were no quantum mechanical communication between the two
classically isolated vortex tubes, the dipole would be exactly
$2T$-periodic. But there is quantum tunneling from one tube through the
stochastic sea to its counterpart, and, as a consequence,
the two corresponding quasienergies do not differ by exactly $ \omega/2 $.
To complete the double well analogy, the deviation from this value has to be
interpreted as the tunnel splitting; in particular, eq.~(\ref{SPL}) describes
the tunnel splitting of the ground state doublet. Its remarkable exponential
suppression with increasing driving strength could have important practical
consequences.

The picture discussed so far can immediately be generalized
to a primary $r$:1 - resonance. In such a case, there are $r$ disconnected
$rT$-periodic vortex tubes, and the Floquet states are extended over all
of them. Appropriate linear combinations, which localize the probability
density along one of these tubes, lead to the generation of subharmonics
with frequency $ \omega/r$. The approximate analytic construction of the
Floquet states then also requires the $r\pi$-periodic Mathieu functions.

{}From a practical point of view, the mechanism of subharmonic generation
outlined in this Letter has several attractive features. It depends only on a
property of the driven potential --- the existence of a 2:1 - resonance ---
but neither on the exact potential shape nor (neglecting the tunneling
corrections) on the exact strength of the driving force.
In addition, {\em{any}} linear superposition of Floquet states that contains
both members of a doublet yields subharmonics.
The particular example of a triangular well potential is realized in
semiconductor heterojunctions~\cite{BAS}, with characteristic energy
spacings in the far-infrared regime. Now that harmonic generation in
far-infrared driven heterostructures has been reported~\cite{SHE},
it is both an experimental and a theoretical challenge to explore
whether our mechanism of subharmonic generation
can be exploited in such mesoscopic devices.

We thank S.J. Allen and M.S. Sherwin for discussions. M.E.F.
was supported by the National Science Foundation under Grant
No. PHY89-04035. M.H. acknowledges support from ONR grant
N00014-92-J-1415 and the Alexander von Humboldt-Stiftung.

\begin{figure}
\caption{Poincar\'e surface of section for the driven triangular
well~(\ref{HAM}) for $ \lambda = 0.4 $ and $\omega = 0.92$,
taken at $ t = 0 $ mod $ T $. }
\end{figure}

\begin{figure}
\caption{ Quasienergies for the driven triangular well with
$\omega=0.92$, as functions of $\lambda$.
	The arrows on the right margin indicate the ground state doublet
	of the 2:1-resonance.}
\end{figure}

\begin{figure}
\caption{Probability density of a member of the ground state doublet
	of the 2:1-resonance, for two periods of the external force.
	Lines connect points of equal density.}
\end{figure}

\begin{figure}
\caption{Probability density of a solution of the time-dependent
	Schr\"{o}dinger equation that  consists of a superposition of both
	members of the ground state doublet. }
\end{figure}

\begin{figure}
\caption{Expectation value of the dipole operator for the wave function
	shown in Fig.~4, for $6T$.}
\end{figure}
\end{document}